\documentclass[conference]{IEEEtran}
\usepackage{cite}
\usepackage{amsmath,amssymb,amsfonts}
\usepackage{algorithmic}
\usepackage{graphicx}
\usepackage{textcomp}
\usepackage{url}
\usepackage{todonotes}
\usepackage{listings}
\usepackage[most]{tcolorbox}
\usepackage{colortbl}
\usepackage{caption}
\usepackage{balance}
\usepackage{tablefootnote}

\lstdefinestyle{yaml}{
    basicstyle=\color{blue}\footnotesize\ttfamily,
    comment=[l]{:},
    commentstyle=\color{black}\ttfamily,
    morecomment=[l]{-},
    numberstyle=\color{gray},    
    breaklines=false,                 
    captionpos=b,                    
    keepspaces=true, 
    numbers=left,
 }

\tcbuselibrary{listingsutf8}
\newtcblisting{codeframe}[1][]{%
    listing only,
    colback=gray!5,
    colframe=black,
    listing options={
        basicstyle=\color{blue}\footnotesize\ttfamily,
        comment=[l]{:},
        commentstyle=\color{black}\ttfamily,
        morecomment=[l]{-},
        numberstyle=\color{gray},    
        breaklines=false,                 
        captionpos=b,                    
        keepspaces=true, 
        numbers=left,
        #1
    },
    left=5pt,
    right=5pt,
    top=5pt,
    bottom=5pt,
    enhanced,
    sharp corners,
    boxrule=0.8pt
}

\begin{document}

\title{Visualizing Cloud-native Applications with KubeDiagrams}

\author{\IEEEauthorblockN{Philippe Merle}
\IEEEauthorblockA{
\textit{Univ. Lille, Inria, CNRS, Centrale Lille, UMR 9189 CRIStAL}\\
F-59000 Lille, France \\
philippe.merle@inria.fr}
\and
\IEEEauthorblockN{Fabio Petrillo}
\IEEEauthorblockA{\textit{École de technologie supérieure - ÉTS Montréal} \\
\textit{Montréal, Canada}\\
fabio.petrillo@etsmtl.ca}
}

\maketitle

\begin{abstract}
Modern distributed applications increasingly rely on cloud-native platforms to abstract the complexity of deployment and scalability. As the de facto orchestration standard, Kubernetes enables this abstraction, but its declarative configuration model makes the architectural understanding difficult. Developers, operators, and architects struggle to form accurate mental models from raw manifests, Helm charts, or cluster state descriptions.
We introduce KubeDiagrams, an open-source tool that transforms Kubernetes manifests into architecture diagrams. By grounding our design in a user-centered study of real-world visualization practices, we identify the specific challenges Kubernetes users face and map these to concrete design requirements. \texttt{KubeDiagrams} integrates seamlessly with standard Kubernetes artifacts, preserves semantic fidelity to core concepts, and supports extensibility and automation. We detail the tool’s architecture, visual encoding strategies, and extensibility mechanisms. Three case studies illustrate how \texttt{KubeDiagrams} enhances system comprehension and supports architectural reasoning in distributed cloud-native systems.
\texttt{KubeDiagrams} addresses concrete pain points in Kubernetes-based DevOps practices and is valued for its automation, clarity, and low-friction integration into real-world tooling environments.
\end{abstract}


\section{Introduction}

Cloud-native applications are engineered to fully exploit modern cloud computing environments by adhering to scalability, elasticity, resilience, and continuous delivery principles. Built as collections of loosely coupled microservices, these applications are typically containerized and orchestrated using platforms like Kubernetes. 
%Cloud-native architectures offer enhanced modularity and portability compared to traditional monolithic systems, enabling rapid development and deployment across distributed infrastructures. This paradigm supports faster release cycles and empowers organizations to adapt quickly to evolving user demands and workloads, establishing cloud-native development as a cornerstone of contemporary software engineering.
Central to this ecosystem, Kubernetes has transformed how cloud-native applications are deployed and managed. By offering a declarative model for infrastructure through YAML manifests, Kubernetes simplifies many operational tasks while introducing its own layer of abstraction and complexity. As systems grow in size and heterogeneity, comprehending the architecture of Kubernetes-managed applications becomes increasingly challenging.

%Remarque 1
DevOps engineers use manual diagramming tools such as Draw.io or Lucidchart to visualize their cluster architectures. However, the fast-paced evolution of microservices and infrastructure in Kubernetes environments makes these static diagrams quickly outdated and error-prone. The gap between live system state and visual documentation often results in confusion, miscommunication, and reduced situational awareness. Moreover, manually navigating YAML manifests, Helm charts, or \texttt{kubectl} outputs across namespaces imposes a high cognitive load and hinders adequate system comprehension.  

In this paper, we introduce \texttt{KubeDiagrams}, an open-source tool that automates the generation of architecture diagrams directly from Kubernetes cluster states or declarative configurations. It bridges the gap between deployment reality and visual documentation, providing a lightweight yet powerful solution for system understanding. 

%Remarque 2.2
\texttt{KubeDiagrams} automates the generation of up-to-date, semantically rich architecture diagrams from live cluster states or declarative configurations such as YAML manifests, Helm charts, and Kustomize files. \texttt{KubeDiagrams} offers a scriptable, low-friction solution that integrates directly into DevOps workflows, continuously aligning system documentation with actual deployment states. It supports over 47 Kubernetes resource kinds (including custom resources) and provides semantic grouping, official iconography, and extensibility, ensuring that diagrams are accurate and cognitively accessible. Ultimately, \texttt{KubeDiagrams} enhance comprehension, reduce maintenance overhead, and bridge the gap between infrastructure-as-code and system understanding.

%Remarque 2.1
A practitioner-centered analysis revealed strong adoption and appreciation of \texttt{KubeDiagrams} in real-world DevOps workflows. Feedback collected from blogs, social media, and technical posts highlights that users value the tool's ability to automate architecture diagram generation, maintain up-to-date documentation, and integrate seamlessly into CI/CD pipelines. Practitioners cited strengths include ease of use, support for infrastructure-as-code, and high visual clarity, reinforcing KubeDiagrams' practical relevance.

The remainder of this paper is structured as follows. 
%Section~\ref{sec:background} introduces the core concepts of Kubernetes that underpin this work. 
Section~\ref{sec:kubediagrams} describes the architecture and functionality of the \texttt{KubeDiagrams} tool. 
Section~\ref{sec:use_cases} illustrates key features through three concrete use cases drawn from real-world scenarios. Section~\ref{sec:compartive} compares \texttt{KubeDiagrams} with existing visualization tools, highlighting differences in scope and implementation. Section~\ref{sec:practitioners} examines practitioner feedback collected from grey literature, offering insight into adoption and perceived value. Section~\ref{sec:limitations} outlines the main limitations of the current tool implementation. Finally, Section~\ref{sec:conclusion} summarizes our findings and outlines directions for future work.

\section{The \texttt{KubeDiagrams} Tool}
\label{sec:kubediagrams}

\texttt{KubeDiagrams} is an open-source software visualization tool designed to generate architecture diagrams of Kubernetes-based systems. Developed as both a command-line utility and a Python library, the tool addresses a key challenge in cloud-native development: keeping architectural documentation accurate, up-to-date, and aligned with infrastructure-as-code (IaC) practices. 

\texttt{KubeDiagrams} automates the transformation of declarative system descriptions (such as Kubernetes manifests, Kustomize overlays, Helm charts, and helmfile descriptors) or live cluster state into clear, structured, and semantically meaningful diagrams.

\subsection{Core Functionality and Features}
\texttt{KubeDiagrams} supports a wide range of input formats and deployment scenarios. It can ingest YAML-based static configurations or connect to a running Kubernetes cluster via \texttt{kubectl}, providing support for both design-time and runtime documentation. The following features highlight its utility and robustness:

\paragraph{Versatile Input Sources}
Users can generate diagrams from:
\begin{itemize}
    \item Live cluster state using \texttt{kubectl get all -o yaml | kube-diagrams -o diagram.png -}
    \item Local YAML manifests representing individual or multi-resource configurations
    \item Kustomize overlays and patched configurations
    \item Helm charts, including remote and OCI-based repositories via the \texttt{helm-diagrams} command
% @Fabio: j'ai ajouté helmfile source
    \item helmfile descriptions composing Helm charts, Kustomize overlays, and Kubernetes manifests together
\end{itemize}

\paragraph{Comprehensive Kubernetes Object Support}
\texttt{KubeDiagrams} supports over 47 Kubernetes resource types including core objects (e.g., \texttt{Pod}, \texttt{Deployment}, \texttt{Service}), network policies, storage and RBAC resources, and Custom Resource Definitions (CRD). The tool ensures complete coverage of both standard and platform-extended workloads, enabling architectural diagrams that reflect production-grade environments.

\paragraph{Semantic Grouping and Label-Based Clustering}
Resources are organized hierarchically based on namespaces and labels (e.g., \texttt{app}, \texttt{app.kubernetes.io/name}, \texttt{tier}). This grouping improves scalability and comprehension by visually delineating application boundaries, services, and system responsibilities.

\paragraph{Graph Semantics and Edge Typing}
Relationships between resources are represented with meaningful edge styles.

\paragraph{Official Kubernetes Iconography}
To reduce cognitive overhead, the tool uses icons from the Kubernetes design language. These visuals improve diagram legibility and align with the mental models of platform engineers and DevOps practitioners familiar with Kubernetes.

\paragraph{Multiple Export Formats}
\texttt{KubeDiagrams} supports output to PNG, JPG, GIF, TIFF, SVG, PDF, and GraphViz DOT, making it suitable for integration into a wide range of documentation and visualization workflows, from markdown-based internal wikis to external presentations and graph analytics tools.

\paragraph{CI/CD and Automation-Friendly}
The tool is lightweight and easily scriptable, available via \texttt{pip} or as a container image. Teams can embed it within CI/CD pipelines like GitHub actions to auto-generate diagrams and publish them to documentation portals, ensuring the system architecture is continuously synchronized with deployments.

\subsection{Visual Semiotics}

\subsubsection{Core Semiotics}

\begin{figure}[htbp]
\centerline{
\includegraphics[width=5cm]{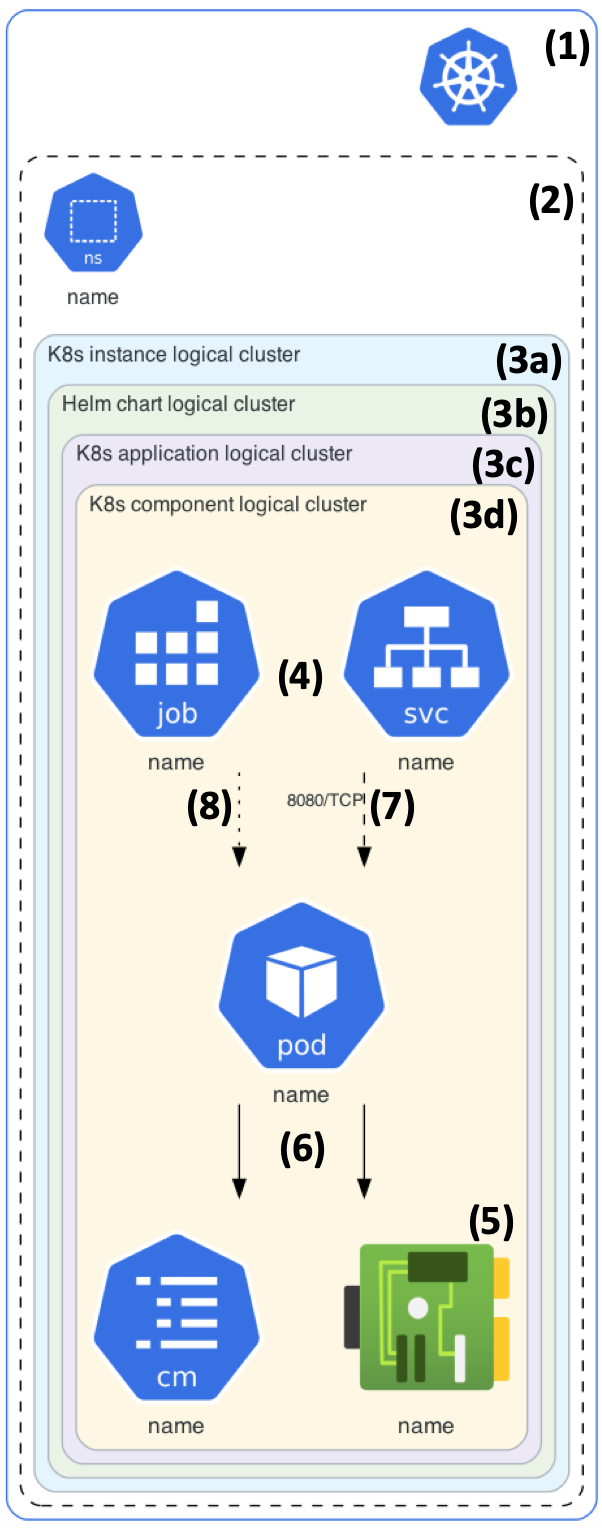}
}
\caption{{KubeDiagrams} visual semiotics.}
\label{fig:visual-semiotics}
\end{figure}

As shown in Fig.~\ref{fig:visual-semiotics}, the KubeDiagrams' visual semiotics is composed of:

\paragraph{Clusters} A visual cluster contains other clusters, resources, and edges. There are two categories of visual clusters:
    \begin{itemize}
    \item System clusters including
        \begin{itemize}
            \item \textbf{(1) Kubernetes cluster} is the top-level cluster containing all the \texttt{namespace}s, logical clusters, and resources composing a running cloud-native system.
            \item \textbf{(2) Namespace cluster} represents a \texttt{Namespace} resource and all its owned/namespaced resources.
        \end{itemize}
    \item Logical clusters including
        \begin{itemize}
        \item \textbf{(3a) K8s instance cluster} contains all the Helm charts, applications, components, resources composing a cloud-native system instance.
        \item \textbf{(3b) Helm chart cluster} contains all the applications, components, resources packaged in a same Helm chart.
        \item \textbf{(3c) K8s application cluster} contains all the components and resources forming a cloud-native application coherently.
        \item \textbf{(3d) K8s component cluster} contains a set of resources forming a coherent part of a cloud-native application.
        \end{itemize}
    \end{itemize}
\paragraph{Nodes} A node is the visual representation of a Kubernetes resource. Its upper part is a visual icon representing the kind of the resource such as \texttt{Job}, \texttt{Service}, \texttt{Pod}, \texttt{ConfigMap}, \texttt{Network\-AttachmentDefinition}. Its lower part is the name of the resource. There are two categories of visual nodes:
    \begin{itemize}
    \item \textbf{(4) Built-in resource} is provided by Kubernetes clusters natively. Its icon respects the iconography defined by the Kubernetes community\footnote{\url{https://github.com/kubernetes/community/tree/master/icons}}. 
    \item \textbf{(5) Custom resource} is not provided by Kubernetes clusters natively such as \texttt{NetworkAttach\-mentDefinition}, \texttt{Certificate}, etc. It requires deploying a custom operator, which defines the structure of the custom resource via a CRD and implements its dynamic behavior via dedicated controllers. The icon of a custom resource is freely defined by operator providers or end-users.
    \end{itemize}
\paragraph{Edges} An edge is the visual representation of a relation between two Kubernetes resources. There are three categories of visual edges:
    \begin{itemize}
    \item \textbf{(6) explicit object reference} (black solid line), \textit{e.g.}, from \texttt{Pod} to \texttt{ConfigMap}
    \item \textbf{(7) label-based selector} (black dashed line), from \texttt{Service} to \texttt{Pod}
    \item \textbf{(8) owner/controller} (black dotted line), \textit{e.g.}, \texttt{Pod} owned by \texttt{Job}
\end{itemize}

Thereby, the KubeDiagrams' visual semiotics is very simple (3 meta-concepts and 8 instantiations) and easily understandable by any Kubernetes practitioner. Moreover, this semiotics is customizable and extensible as illustrated later in Section \ref{sec:use_cases}.

\subsubsection{Supported Resource Types and Icons}

{KubeDiagrams} supports visualization of core and extended Kubernetes resources including but not limited to:
\begin{itemize}
  \item Workloads: \texttt{Pod}, \texttt{Deployment}, \texttt{StatefulSet}, \texttt{DaemonSet}, \texttt{Job}, \texttt{CronJob}, \texttt{Replication\-Controller}, \texttt{PodTemplate}
  \item Configuration: \texttt{ConfigMap}, \texttt{Secret}
  \item Scaling: \texttt{HorizontalPodAutoscaler}, \texttt{Vertical\-PodAutoscaler}
  \item Policies: \texttt{LimitRange}, \texttt{PodDisruptionBudget}, \texttt{PodSecurityPolicy}, \texttt{ResourceQuota}
  \item Network: \texttt{Service}, \texttt{Endpoints}, \texttt{EndpointSlice}, \texttt{Ingress}, \texttt{IngressClass}, \texttt{NetworkPolicy}, \texttt{NetworkAttachmentDefinition}
  \item Storage: \texttt{PersistentVolume}, \texttt{PersistentVolu\-meClaim}, \texttt{StorageClass}, \texttt{VolumeAttachment}, \texttt{CSINode}, \texttt{CSIDriver}, \texttt{CSIStorageCapacity}
  \item RBAC: \texttt{ServiceAccount}, \texttt{Role}, \texttt{RoleBinding}, \texttt{ClusterRole}, \texttt{ClusterRoleBinding}, \texttt{User}, \texttt{Group}
  \item Control Plane: \texttt{Node}, \texttt{PriorityClass}, \texttt{Runtime\-Class}, \texttt{APIService}
  \item Custom resources: \texttt{CustomResourceDefinition},
  \texttt{ValidatingWebhookConfiguration},
  \texttt{Muta\-ting\-WebhookConfiguration}
\end{itemize}

Mappings between these resource kinds and visual elements (nodes, edges, clusters) are defined in an internal configuration file and can be extended through users' custom configurations.

\subsubsection{Extensibility and Customization}
\texttt{KubeDiagrams} can be customized through external configuration files written in YAML. Users can define:
\begin{itemize}
    \item Custom visual mappings for CRDs or extended resource types
    \item Logical clusters to group heterogeneous resources (e.g., external services, legacy systems)
    \item Additional node and edge types for domain-specific semantics
\end{itemize}
This extensibility enables adaptation to varied deployment topologies and supports domain-specific visualizations beyond the default Kubernetes model.

\subsection{Installation and Integration}
Installation is straightforward:
\begin{verbatim}
pip install KubeDiagrams
\end{verbatim}
or via Docker:
\begin{verbatim}
docker run -v "$(pwd)":/work \
philippemerle/KubeDiagrams kube-diagrams \
-o output.png input.yaml
\end{verbatim}

These options enable flexible integration into local development environments or automated workflows. For example, documentation repositories can be configured to regenerate diagrams on every Git push or cluster deployment.

\subsection{Usage Scenarios}
The tool has been applied in a variety of real-world software engineering contexts:
\begin{itemize}
    \item \textbf{Continuous Architecture Documentation:} Ensures that architecture diagrams reflect the current system state in CI/CD pipelines.
    \item \textbf{Multi-Environment Comparison:} Highlights configuration drift across environments (e.g., dev, staging, production).
    \item \textbf{System Comprehension and Onboarding:} Facilitates faster onboarding and better system understanding through visual explanations.
    \item \textbf{Operational Analysis and Debugging:} Diagrams generated during incident response can help identify dependencies and potential points of failure.
    \item \textbf{Architecture Review and Compliance Audits:} Offers visual artifacts for validating system design and audit reporting.
\end{itemize}

\subsection{Implementation}

% The main components are as follows:

% \begin{itemize}
%   \item \textbf{Input Handler:} Supports input from standard input (stdin), local files, Helm templates, or direct cluster state via \texttt{kubectl}. It detects input type and routes parsing accordingly.
%   \item \textbf{Parser and Mapper:} Parses Kubernetes resource definitions using \texttt{PyYAML} and maps them to semantic diagram elements. The mapping logic includes handling of default Kubernetes resources as well as user-defined Custom Resource Definitions (CRDs).
%   \item \textbf{Visualizer:} Leverages the \texttt{diagrams} library to render architecture graphs. Each Kubernetes object (e.g., Pod, Service, Deployment) is mapped to a visual node, with customizable icons and layout logic. Resources can be grouped into clusters to reflect namespaces or logical groupings.
%   \item \textbf{Configuration Layer:} The tool can be configured via a user-provided \texttt{KubeDiagrams.yaml} file, enabling custom resource mappings, filtering, and layout preferences.
% \end{itemize}

\begin{figure}[htbp]
\centerline{
\includegraphics[width=\linewidth]{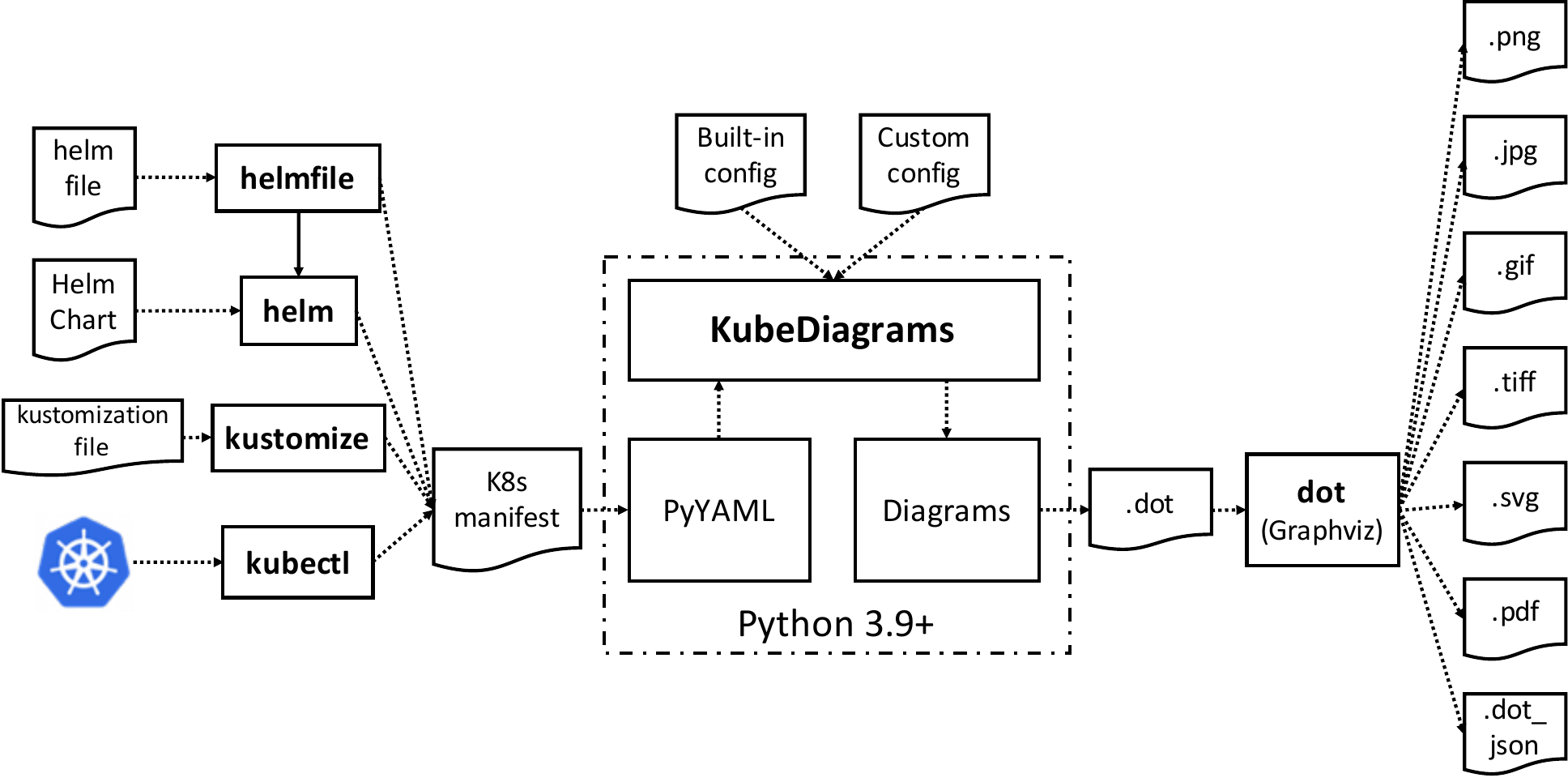}
}
\caption{\texttt{KubeDiagrams} software architecture.}
\label{fig:architecture}
\end{figure}

Fig.~\ref{fig:architecture} illustrates the high-level software architecture and processing pipeline of KubeDiagrams. At the input stage, Kubernetes resources can be sourced from either helmfile descriptors (via the \texttt{helmfile} command) or Helm charts (via the \texttt{helm} command) or kustomization files (via the \texttt{kustomize} command) or directly from live clusters (via the \texttt{kubectl} command). These sources produce Kubernetes manifests in YAML format, which serve as the primary input for the tool.

KubeDiagrams, implemented in Python 3.9+, parses these manifests using the PyYAML library to extract structured data. Next, \texttt{KubeDiagrams} transforms this data into a visual representation. The tool is configurable through both built-in and custom configuration files, allowing users to customize visual mappings, such as icons, labels, and groupings, including support for custom resources (CRD).

Once the internal representation is constructed, \texttt{KubeDiagrams} outputs a .dot file, the standard format used by Graphviz \cite{10.1007/3-540-45848-4_57} for describing graphs. This file is then processed by Graphviz's \texttt{dot} utility to generate visual outputs in multiple formats, including PNG, JPG, GIF, TIFF, SVG, PDF, to name a few.

Overall, \texttt{KubeDiagrams} acts as a pipeline that transforms declarative infrastructure-as-code (IaC) into expressive, customizable architecture diagrams, supporting automation, extensibility, and multiple output formats suitable for documentation or presentation.

\texttt{KubeDiagrams} is distributed as a Python package\footnote{\url{https://pypi.org/project/KubeDiagrams}}, a container image for running \texttt{KubeDiagrams} inside a container\footnote{\url{https://hub.docker.com/r/philippemerle/kubediagrams}}, a Nix flake for reproducible builds\footnote{\url{https://github.com/philippemerle/KubeDiagrams/blob/main/flake.nix}}, and a GitHub action for generating diagrams at CI/CD time\footnote{\url{https://github.com/philippemerle/KubeDiagrams/blob/main/action.yml}}.

\subsection{Availability}

\texttt{KubeDiagrams} is available under the GPL-3.0 license and can be accessed via GitHub at:
\begin{center}
\url{https://github.com/philippemerle/KubeDiagrams}
\end{center}
Its open-source nature, low installation barrier, and strong alignment with DevOps workflows make it a valuable addition to the software visualization toolbox for modern Kubernetes-based systems.

\section{Use Cases}
\label{sec:use_cases}

This section illustrates three use cases of KubeDiagrams. \texttt{KubeDiagrams} could be applied to 1) any Kubernetes-based business application, \textit{e.g.}, the well-known \texttt{WordPress} publishing platform, 2) any Kubernetes operator implementing controllers for custom resources, \textit{e.g.}, the well-known \texttt{cert-manager} operator, and 3) any Kubernetes control plane implementation, \textit{e.g.}, the well-known \texttt{minikube} one.

\subsection{\texttt{WordPress} Publishing Platform}

\begin{figure}[htbp]
\centerline{
\includegraphics[width=.80\linewidth]{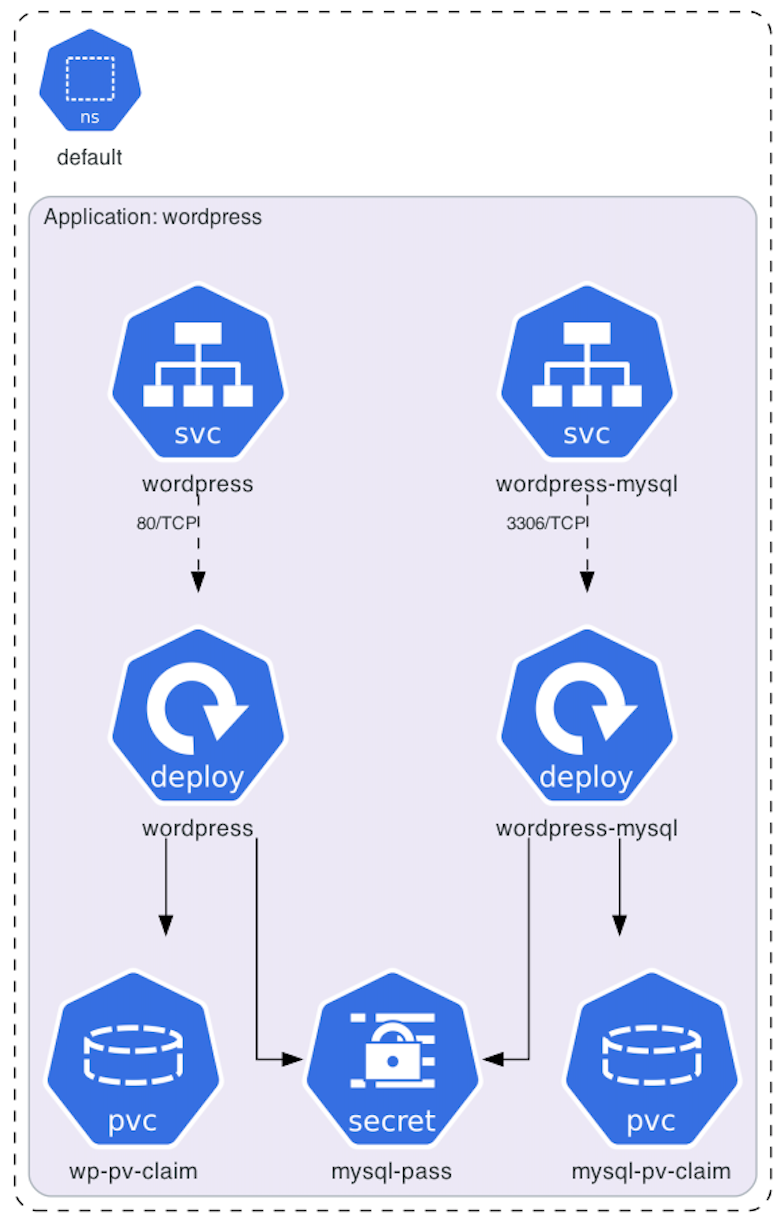}
}
\caption{Generated diagram for \texttt{WordPress} manifests.}
\label{fig:wordpress}
\end{figure}

One of the official Kubernetes tutorials\footnote{\url{https://kubernetes.io/docs/tutorials/stateful-application}} is based on \texttt{WordPress}, an open source publishing platform used by millions of websites worldwide. 
Fig.~\ref{fig:wordpress} shows the diagram generated by \texttt{KubeDiagrams} automatically from the tutorial's manifests. As shown by the diagram,
this use case is composed of two deployment workloads: \texttt{wordpress} encapsulates the publishing platform and \texttt{wordpress-mysql} encapsulates a MySQL database manager. Each workload is exposed by its own service (\texttt{svc}) and has its own persistent volume claim (\texttt{pvc}). A \texttt{secret} containing the password to access the database is shared between both workloads.
Dashed black arrows represent selectors from services to workloads and solid black arrows represent explicit references between resources.

All these resources are in the \texttt{default} namespace represented by the dashed black container.
As they are all labelled with the same label (\texttt{app: wordpress}) then all of them are grouped in the light purple container. The mapping between the label \texttt{app} and the visual container attributes (\texttt{title} and  \texttt{bgcolor}) is defined in the following \texttt{KubeDiagrams} custom configuration:

\begin{minipage}{0.45\textwidth}
\begin{lstlisting}[frame=single, style=yaml]
clusters:
  - label: app
    title: "Application: {}"
    bgcolor: "#ECE8F6"
\end{lstlisting}
\end{minipage}

\begin{figure}[htbp]
\centerline{
\includegraphics[width=0.85\linewidth]{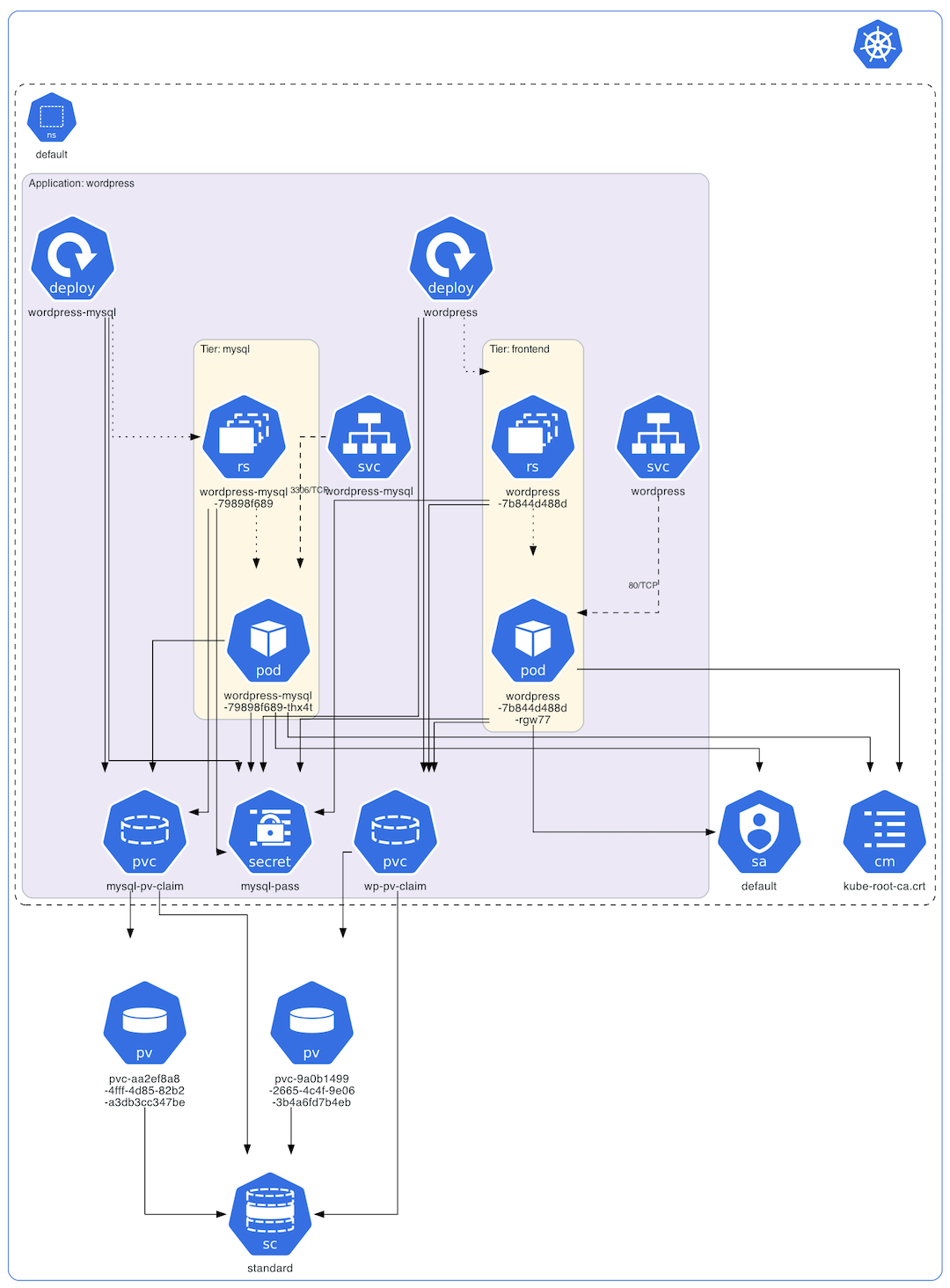}
}
\caption{Generated diagram for a deployed \texttt{WordPress} instance.}
\label{fig:deployed_wordpress}
\end{figure}

Fig.~\ref{fig:deployed_wordpress} shows a generated diagram representing a \texttt{WordPress} instance deployed on a Kubernetes cluster represented by the encompassing blue container. The deployment of both \texttt{WordPress} workloads instantiates new resources compared to Fig.~\ref{fig:wordpress}, \textit{i.e.}, each deployment owns a replica set (\texttt{rs}) managing \texttt{pod}s, the \texttt{default} namespace owns a service account (\texttt{sa}) and a config map (\texttt{cm}) containing Kubernetes credentials, and each persistent volume claim owns a persistent volume (\texttt{pv}) of the \texttt{standard} storage class (\texttt{sc}).

\smallskip

This use case shows that developers could use \texttt{KubeDiagrams} to visualize their business cloud-native applications (here \texttt{WordPress}) at both development (Fig.~\ref{fig:wordpress}) and deployment (Fig.~\ref{fig:deployed_wordpress}) times with minimal cognitive efforts.

\subsection{\texttt{cert-manager} Operator}

\texttt{cert-manager}\footnote{\url{https://cert-manager.io}} is the world-wide leading cloud native X.509 certificate management. Concretely, \texttt{cert-manager} is a Kubernetes operator controlling both \texttt{Issuer} and \texttt{Certificate} custom resources. 

As declared in Listing \ref{lst:issuer-certificate}, a certificate is managed by a certificate authority called issuer (Lines 13-14) and is stored with its signed private key in a dynamically created secret (Line 15). \texttt{cert-manager} deals with a variety of certificate authorities, as the self-signed issuer declared in Line 6. 

\noindent
\lstset{numbersep=4pt}
\hspace{0.2cm}
\begin{minipage}{0.27\textwidth}
%\begin{codeframe}[
\begin{lstlisting}[xleftmargin=-0.8ex, frame=single, style=yaml,
caption=Declaration of \texttt{Issuer} and \texttt{Certificate} custom resources.,label={lst:issuer-certificate}]
apiVersion: cert-manager.io/v1
kind: Issuer
metadata:
  name: selfsigned-issuer
spec:
  selfSigned: {}
---
apiVersion: cert-manager.io/v1
kind: Certificate
metadata:
  name: serving-cert
spec:
  issuerRef:
    name: selfsigned-issuer
  secretName: serving-cert
\end{lstlisting}
%\end{codeframe}
\end{minipage}
\hfill
\begin{minipage}{0.17\textwidth}
  \centering
  \includegraphics[width=1.05cm]{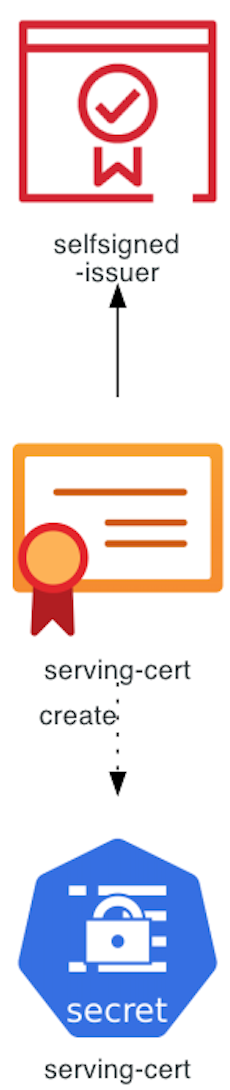}
  \captionof{figure}{Visual representation of \texttt{Issuer} and \texttt{Certificate} custom resources.}
    \label{fig:certificate-issuer}
\end{minipage}

\begin{figure*}[htbp]
\centerline{
\includegraphics[width=0.83\linewidth]{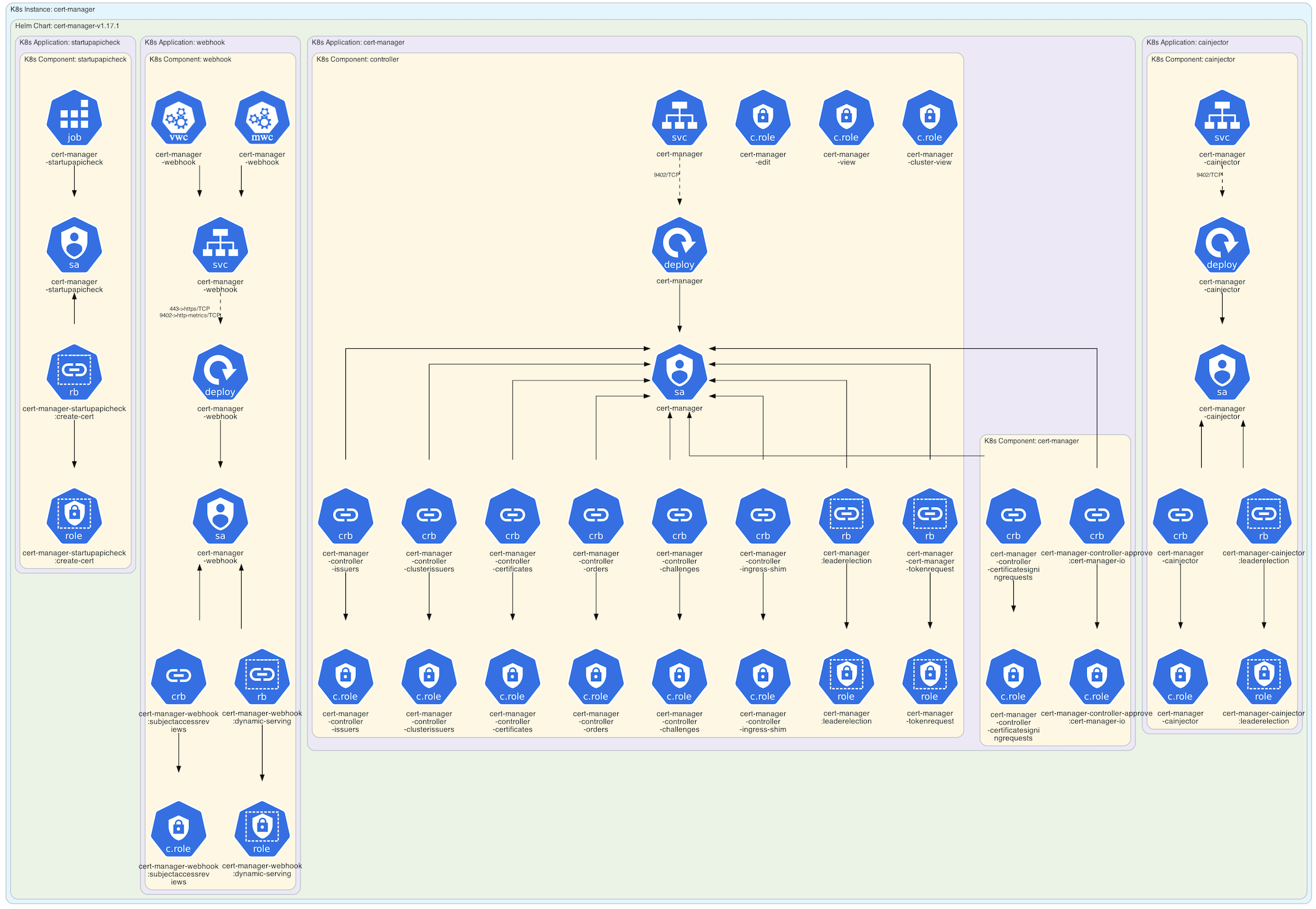}
}
\caption{Generated diagram for \texttt{cert-manager} Helm Chart.}
\label{fig:cert-manager}
\end{figure*}

\texttt{KubeDiagrams} allows one to associate a visual representation to any Kubernetes custom resource as illustrated in Fig.~\ref{fig:certificate-issuer}. The mapping between a custom resource type and its visual representation is declaratively defined as shown in Listing \ref{lst:issuer-certificate-config}.
Custom resource types are identified by the concatenation of their \texttt{kind} and \texttt{apiVersion} (Lines 2 and 5). The scope of a custom resource type is either \texttt{Namespaced} or \texttt{Cluster} (Lines 3 and 6). A visual icon is associated to each custom resource type (Lines 4 and 7). A custom resource could dynamically create other resource as illustrated in Lines 8-11. The reference and selector fields of a custom resource (Lines 13 and 20) are mapped to visual edge attributes (Lines 16-19 and 21-24).

\begin{minipage}{0.45\textwidth}
\begin{lstlisting}[frame=single, style=yaml, caption=\texttt{KubeDiagrams} custom configuration for \texttt{cert-manager}.,label={lst:issuer-certificate-config}]
nodes:
  Issuer/cert-manager.io/v1:
    scope: Namespaced
    custom_icon: issuer.png
  Certificate/cert-manager.io/v1:
    scope: Namespaced
    custom_icon: certificate.png
    nodes:
      spec.secretName:
        kind: Secret
        apiVersion: v1
    edges:
      spec.issuerRef.name:
        kind: Issuer
        apiVersion: cert-manager.io/v1
        graph_attr:
          color: black
          style: solid
          direction: up
      spec.secretName:
        graph_attr:
          xlabel: create
          color: black
          style: dotted
\end{lstlisting}
\end{minipage}

Fig.~\ref{fig:cert-manager} shows the generated diagram for the Helm chart
packaging \texttt{cert-manager}\footnote{\url{https://artifacthub.io/packages/helm/cert-manager/cert-manager}}. Let us note that this generated diagram clearly summarizes key architectural aspects encoded in around 1.444 lines of YAML manifests. The colored containers represent the hierarchical structure of \texttt{cert-manager} including its Helm charts (here just one), applications, components, and resources. Let us note that most of its resources are dedicated to the configuration of Kubernetes role-based access control policies: roles (\texttt{role} and \texttt{c.role}), role bindings (\texttt{rb} and \texttt{crb}), and service accounts (\texttt{sa}). Two resources (\texttt{vwc} and \texttt{mwc}) extend the Kubernetes API server in order to validate and mutate \texttt{Certificate} and \texttt{Issuer} custom resources. Finally, four workload resources (\texttt{deploy} and \texttt{job}) encapsulate the functional code of \texttt{cert-manager}.

\smallskip

This use case shows how \texttt{KubeDiagrams} helps to visualize any cloud-native operator (here \texttt{cert-manager}) and its controlled custom resources (here \texttt{Certificate} and \texttt{Issuer}) with few cognitive efforts.

\begin{figure}[htb]
\centerline{
\includegraphics[width=\linewidth]{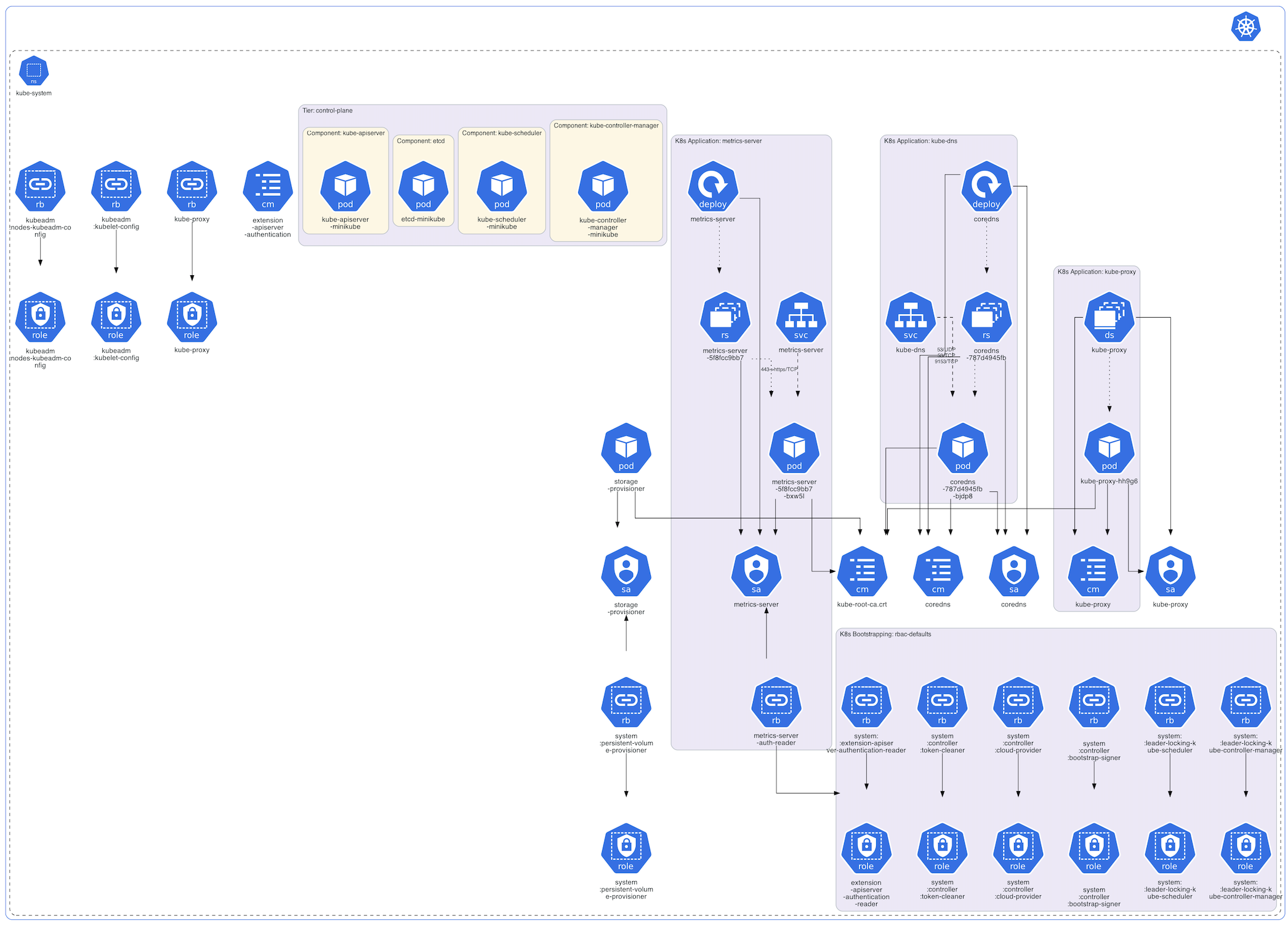}
}
\caption{Generated diagram for \texttt{minikube} control plane.}
\label{fig:minikube-control-plan}
\end{figure}

\subsection{\texttt{minikube} control plane}

\texttt{minikube}\footnote{\url{https://minikube.sigs.k8s.io}} quickly sets up a local Kubernetes cluster on macOS, Linux, Windows, and it is mainly used by developers rather than in production. As shown in Fig.~\ref{fig:minikube-control-plan}, the \texttt{kube-system} namespace contains 1) the \texttt{control-plane} and its \texttt{kube-apiserver}, \texttt{etcd}, \texttt{kube-scheduler}, and \texttt{kube-controller-manager} components, 2) a network communication service (\texttt{kube-proxy}), 3) a local DNS service (\texttt{kube-dns}), 4) a local storage service, 5) a metrics service (\texttt{metrics-server}), and 6) bootstrapping role-based access control policies.

The  three previous use cases show that \texttt{KubeDiagrams} is helpful to visualize, document, and understand all the layers of any cloud-native system including the Kubernetes control plane (\textit{e.g.}, \texttt{minikube}), custom controllers (\textit{e.g.}, \texttt{cert-manager}), and business applications (\textit{e.g.}, \texttt{WordPress}).
The KubeDiagrams's public repository contains many other use cases\footnote{\url{https://github.com/philippemerle/KubeDiagrams/tree/main/examples}} such as the \texttt{k0s} lightweight Kubernetes control plane for embedded systems, several 5G core network functions, various custom controllers, and business applications.

\section{Tools' comparative analysis}
\label{sec:compartive}

\begin{table*}[ht]
\centering
\caption{Comparison of Kubernetes Visualization Tools - Activity and Popularity} 
\label{tab:activity}
\begin{tabular}{lllccc}
\hline
\textbf{Tool} & \textbf{Created At} & \textbf{Last Commit} & \textbf{Contributors} & \textbf{Forks} & \textbf{Stars} \\
\hline
\rowcolor{lightgray}KubeView \cite{kubeview} & February 2019 & November 2022 & 5 & 116 & 993\\
\textbf{KubeDiagrams} \cite{kubediagrams}& \textbf{December 2024} & \textbf{May 2025} & \textbf{8} & \textbf{32} & \textbf{803}\\
\rowcolor{lightgray}k8sviz \cite{k8sviz}&  August 2019 & June 2022 & 2 & 51 & 309 \\
Kubernetes diagrams \cite{k8sdiagrams2021}  & April 2021 & February 2024 & 3 & 12 & 143 \\
\rowcolor{lightgray}GruCloud \cite{grucloud}  & October 2020 & September 2024 & 4 & 14 & 118 \\
k8s-to-diagram \cite{kocierik2024k8sdiagram}  & July 2024 & September 2024 & 2 & 2 & 20 \\
\rowcolor{lightgray}react-k8s-viewer \cite{reactk8sviewer} & January 2021 & November 2024 & 5 & 1 & 13 \\
K8s Diagram Previewer \cite{k8sdiagrampreviewer} & May 2021 & July 2021 & 1 & 1 & 8 \\
\rowcolor{lightgray}k8s-diagrams \cite{k8s-diagrams} & October 2021 & October 2021 & 1 & 1 & 2 \\
kube-diagram \cite{kube-diagram} & January 2022 & December 2024 & 1 & 0 & 1 \\
\rowcolor{lightgray}kube-diagrams \cite{kubeDiagrams2020} & November 2020 & April 2023 & 1 & 0 & 1 \\
k8d \cite{k8d}& January 2020 & January 2020 & 1 & 0 & 1 \\
\rowcolor{lightgray}k8s\_diagram \cite{k8s_diagram2024}& April 2024 & May 2024 & 1 & 0 & 0 \\
KubeDraw \cite{kubedraw2024} & May 2024 & May 2024 & 1 & 0 & 0 \\
\hline
\end{tabular}
\end{table*}

\begin{table*}[ht]
\centering
\caption{Comparison of Kubernetes Visualization Tools - Features}

\label{tab:features}
\begin{tabular}{p{3.1cm}cp{2.9cm}p{3.7cm}ccp{2.7cm}}
\hline
\textbf{Tool} & \textbf{Kinds} & \textbf{Implementation} & \textbf{Input formats} & \textbf{KIS$^a$} & \textbf{Cluster depth}& \textbf{Output formats} \\
\hline
\rowcolor{lightgray}KubeView \cite{kubeview} & 10 &  Vue and Go & K8s API & Supported & 0 & Web pages\\
\textbf{KubeDiagrams} \cite{kubediagrams}& 47 & Python with Diagrams & K8s manifests, kustomization files, Helm charts, and K8s API & Supported & 6 & PNG, JPG, GIF, TIFF, SVG, PDF, and DOT\\
\rowcolor{lightgray}k8sviz \cite{k8sviz}& 12 & Go and Graphviz & K8s API & Supported & 1 & All Graphviz outputs\\
Kubernetes diagrams \cite{k8sdiagrams2021} & 8 & Go & K8s API & Supported & 2 & DOT \\
\rowcolor{lightgray}GruCloud \cite{grucloud} & 14 & Javascript & Javascript IaC & Unsupported & 3 & PlantUML \\
k8s-to-diagram \cite{kocierik2024k8sdiagram} & 9 & Go and D2 & K8s manifest annotations& Unsupported & 2 & SVG and PNG \\
\rowcolor{lightgray}react-k8s-viewer \cite{reactk8sviewer} & 7 & Typescript & K8s manifests & Unsupported & 0 & React Flow \\
K8s Diagram Previewer \cite{k8sdiagrampreviewer} & 11 & Python with Diagrams & K8s manifests and Helm charts & Supported & 2 & PNG, JPG, SVG, PDF, and DOT \\
\rowcolor{lightgray}k8s-diagrams \cite{k8s-diagrams} & 8 & Python with Diagrams & K8s API & Supported & 1 & PNG, JPG, SVG, PDF, and DOT \\
kube-diagram \cite{kube-diagram} & 5 & Java & K8s manifests & Unknown & Unknown & Unknown \\
\rowcolor{lightgray}kube-diagrams \cite{kubeDiagrams2020} & 3 & Python with Diagrams & K8s API & Supported & 2 & PNG \\
k8d \cite{k8d}& 2 & Go & K8s API & Unknown & Unknown & XML for draw.io \\
\rowcolor{lightgray}k8s\_diagram \cite{k8s_diagram2024}& 8 & Python & K8s API & Supported & Unknown & PNG \\
KubeDraw \cite{kubedraw2024} & 5 & Python with Diagrams & K8s API & Supported & 0 & PNG \\
\hline
\scriptsize{$^a$Kubernetes Icons Set (KIS)}
\end{tabular}
\end{table*}

The Kubernetes visualization tools landscape demonstrates considerable variation in scope, activity, and adoption. \texttt{KubeDiagrams} leads in terms of Kubernetes resource coverage, supporting 47 distinct Kubernetes resource kinds—more than four times that of many tools. Its development also remains current, with the last commit dated May 2025 and eight contributors maintaining the project. The high engagement correlates with strong popularity, as evidenced by 803 GitHub stars.

KubeView \cite{kubeview}, while older—created in February 2019—still maintains the largest user base, with 993 stars. Despite its last commit occurring in 2022, the tool has garnered attention for its stability and mature features. However, its support covers only 10 resource kinds, which limits its use for more complex cloud-native systems.

In contrast, newer tools like K8s Diagram Architecture Generator \cite{kocierik2024k8sdiagram} and KubeDraw \cite{kubedraw2024} emerged in 2024. Both show recent activity and initial community interest, but their GitHub stars remain modest (20 and 0 respectively). They each list only one or two contributors, which may pose risks for long-term maintenance.

Projects like k8sviz \cite{k8sviz} and Kubernetes diagrams \cite{k8sdiagrams2021} sit in the middle ground. They offer moderate resource kind support (12 and 8 respectively), have received updates within the last few years, and retain small but active contributor bases. These tools may serve well for focused use cases or educational purposes but lack the broader ecosystem coverage seen in KubeDiagrams. Several other tools, including k8s-diagrams \cite{k8s-diagrams}, kube-diagram \cite{kube-diagram}, and k8d \cite{k8d}, have not seen updates since their creation, and each has only a single contributor. These attributes suggest limited ongoing support and a lower likelihood of future enhancements.

The Kubernetes visualization ecosystem significantly varies in implementation strategies, input handling, and output capabilities. Among the surveyed tools (Table \ref{tab:features}), only \texttt{KubeDiagrams} supports the full range of input sources—including raw manifests, Helm charts, Kustomize files, and live cluster state. Most other tools limit input to the Kubernetes API or require annotations, reducing flexibility in Infrastructure-as-Code workflows. Tools such as k8sviz \cite{k8sviz}, Kubernetes diagrams \cite{k8sdiagrams2021}, and k8s-diagrams \cite{k8s-diagrams} consume live cluster data but offer fewer customization options. Regarding implementation, Go and Python dominate, with several tools leveraging the Diagrams library for static rendering. However, only a few tools are exported to multiple formats. \texttt{KubeDiagrams} supports seven output types (PNG, JPG, GIF, TIFF, SVG, PDF, DOT), making it more suitable for integrating diverse documentation workflows. Other tools like GruCloud \cite{grucloud} or k8d \cite{k8d} restrict output to single-purpose formats such as PlantUML or draw.io XML, which may limit reuse. Some tools, including react-k8s-viewer \cite{reactk8sviewer}, focus on web-based rendering but sacrifice portability in the documentation. Only \texttt{KubeDiagrams} combines wide Kubernetes kind coverage (47), extensive input compatibility, and broad output format support. These attributes suggest it fills a unique position in the tooling landscape, balancing completeness, automation, and integration.

\begin{figure}[ht]
\includegraphics[width=\linewidth]{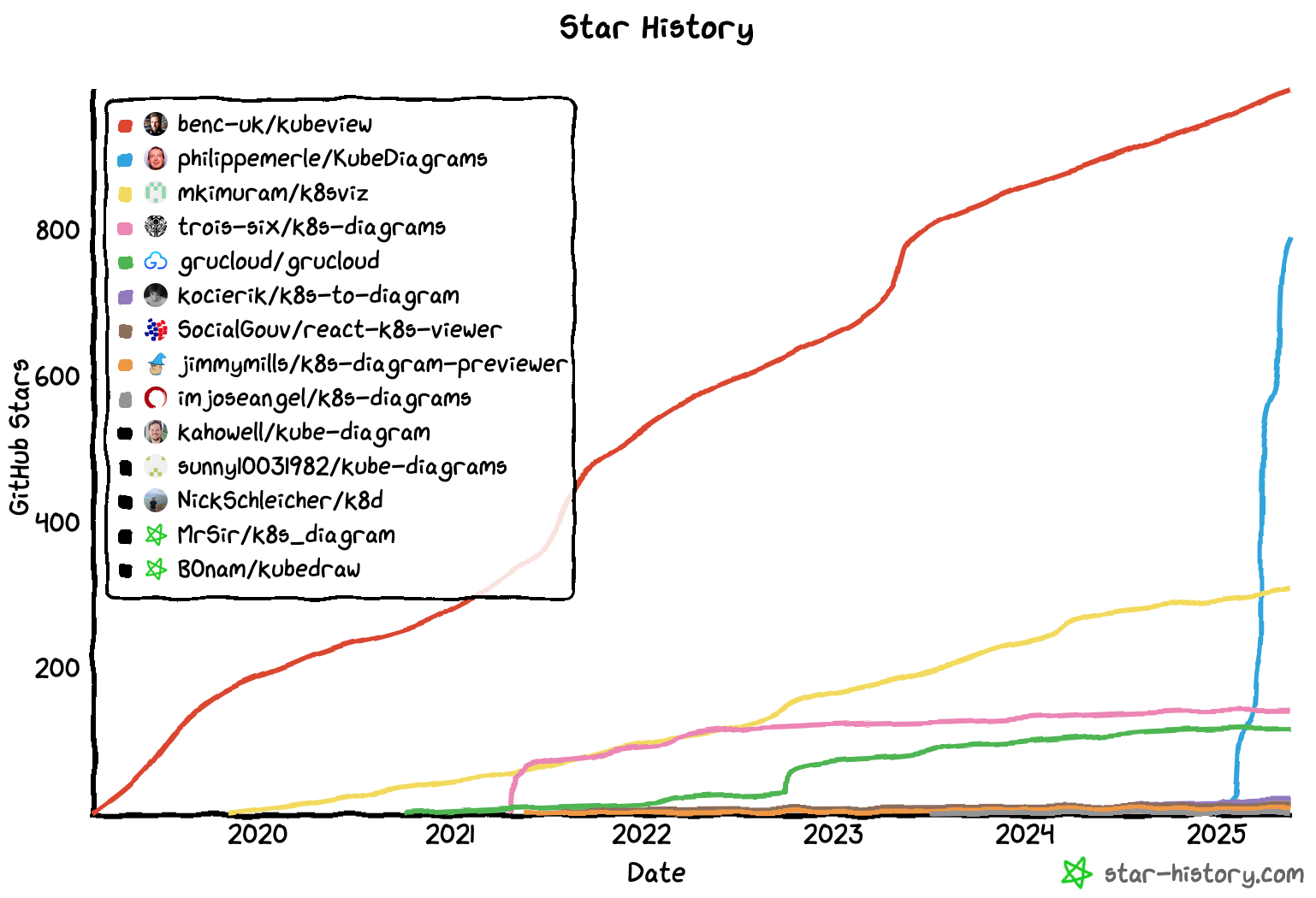}
\caption{Star history of diagrams generators.}
\label{fig:popularity}
\end{figure}

To contextualize the relevance and adoption of \texttt{KubeDiagrams} within the broader ecosystem of Kubernetes visualization tools, we conducted a comparative analysis using GitHub star history as a proxy for community interest and adoption. Fig.~\ref{fig:popularity} illustrates the evolution of GitHub stars over time for fifteen representative projects.
The most prominent project in terms of long-term popularity is \texttt{ben-cuk/kubeview}, which has shown steady growth since 2019 and near 1000 GitHub stars by early 2025. It suggests sustained interest due to early market entry and consistent maintenance. In contrast, \texttt{KubeDiagrams} has demonstrated exceptional recent growth, rapidly accumulating 803 stars in less than six months. This steep curve indicates an accelerating adoption trend, positioning \texttt{KubeDiagrams} as a rising contender and potentially the most rapidly growing tool in this space. 

Furthermore, \texttt{mkimuram/k8sviz} and \texttt{trois-six/ k8s-diagrams} represent stable, moderately adopted alternatives, both steadily accumulating 309 and 143 stars, respectively. Their curves indicate slow but consistent community engagement, likely due to mature feature sets but less recent innovation.
Several other tools show niche or stagnant adoption trajectories, such as \texttt{grucloud/grucloud}, \texttt{kocierik/k8s-to-diagram}, and \texttt{SocialGouv/react-k8s-viewer} display modest growth under 200 stars, with nearly flat trends since 2022.

Overall, Fig.~\ref{fig:popularity} reveals a bifurcation in the Kubernetes visualization ecosystem. Tools like kubeview, k8sviz, and k8s-diagrams reflect steady-state maturity (2 years without contributions), while \texttt{KubeDiagrams} exemplifies a rapid adoption in an active tool project. \texttt{KubeDiagrams} appears the most actively developed and comprehensive option, while KubeView retains a lead in community adoption. Other tools fill niche roles or remain dormant. Researchers and practitioners should weigh functionality, development activity, and contributor engagement when selecting a visualization tool for Kubernetes.
Up-to-date Table~\ref{tab:activity}, Table~\ref{tab:features}, and Fig.~\ref{fig:popularity} data are available online\footnote{\url{https://github.com/philippemerle/Awesome-Kubernetes-Architecture-Diag} \url{rams#generation-tools}}.

\section{Practitioners' Perspective}
\label{sec:practitioners}

To understand how practitioners perceive \texttt{KubeDiagrams} in real-world settings, we systematically searched for unprompted user feedback across several public platforms, including Reddit, Twitter, Medium, and personal technical blogs\footnote{A complete up-to-date list is available at \url{https://github.com/philippemerle/KubeDiagrams#what-do-they-say-about-it}}. Our goal was not to rely solely on curated testimonials or official documentation, but rather to gather authentic, self-initiated observations from software engineers and DevOps professionals who have used the tool in practice.

This exploratory search yielded 6 distinct sources containing explicit commentary on KubeDiagrams. These documents included blog posts detailing integration experiences, Reddit threads discussing usability trade-offs, and social media posts highlighting strengths and limitations. We treated each artifact as a data point reflecting spontaneous practitioner engagement. Through close reading and thematic analysis, we extracted common patterns, concerns, and endorsements from these sources. These insights allowed us to identify the tool's position within current DevOps workflows and how its perceived value compares to other visualization alternatives in the Kubernetes ecosystem.

Abhimanyu Saharan, a DevOps engineer, documents in a blog post ``\textit{Generate Kubernetes Architecture Maps Directly from Your Cluster}" \cite{saharan2025kubediagrams} his adoption of \texttt{KubeDiagrams} to address persistent challenges with maintaining accurate Kubernetes architecture documentation. He identifies a recurring issue in production environments: architecture diagrams often fall out of sync with actual deployments, creating confusion during onboarding, troubleshooting, and audits. Traditional diagramming tools, such as Lucidchart and Draw.io, require manual updates and do not scale with the dynamic nature of Kubernetes. 

Saharan integrates \texttt{KubeDiagrams} into his workflow to automate this task. He highlights the tool’s ability to generate up-to-date architecture diagrams directly from a live Kubernetes cluster or manifests, Helm charts, and Kustomize configurations. KubeDiagrams’ CLI and support for a wide range of Kubernetes resource types (including CRDs) make it immediately valuable for operational and pre-deployment contexts.

The blog emphasizes specific benefits observed in practice: improved environment consistency across dev, staging, and prod; clearer onboarding through live diagrams; and faster incident response via visual system maps. Saharan also demonstrates how \texttt{KubeDiagrams} fits naturally into CI/CD pipelines, enabling continuous documentation generation with minimal configuration. He concludes that \texttt{KubeDiagrams} eliminates the lag between system state and system documentation, replacing guesswork with precise, always-current diagrams. Saharan explicitly states that : \textit{``It solved the ‘outdated diagram’ problem in one swoop by always reflecting the live environment."}; \textit{``By combining real-time data with smart grouping and rich support for Kubernetes resources, \texttt{KubeDiagrams} delivers diagrams that are both accurate and instantly informative."}.
% @Fabio déjà dit avant
% ``By combining real-time data with smart grouping and rich support for Kubernetes resources, \texttt{KubeDiagrams} delivers diagrams that are both accurate and instantly informative."

A practitioner blog post by Mr.PlanB \cite{mrplanb2025kubediagrams} highlights the usability and practical value of \texttt{KubeDiagrams} in everyday DevOps tasks. The author describes the challenge of understanding Kubernetes architectures scattered across YAML files, Helm charts, and multiple namespaces. \texttt{KubeDiagrams} addresses this problem by producing clean architecture diagrams directly from live clusters or configuration files using simple command-line invocations. The author emphasizes the tool’s minimal setup (no UI, no configuration overhead) and notes that it integrates well into pipelines and automation scripts. The post underlines a recurring theme: the tool reduces cognitive and operational overhead without overcomplicating the workflow. The blogger explicitly states that : \textit{“You run a command, and out pops a PNG showing how everything connects. Simple.”}; \textit{“It’s also helping folks troubleshoot and document their setups more easily.”}; \textit{“For anyone who’s spent too long trying to draw a Kubernetes diagram manually, that’s a pretty big win.”}.

A technical blog by dbafromthecold \cite{dbafromthecold2025sql} demonstrates the application of \texttt{KubeDiagrams} to a Kubernetes deployment of Microsoft SQL Server using StatefulSets and persistent volumes. The author emphasizes the complexity of visualizing stateful workloads and highlights the difficulty of maintaining clear documentation across multiple YAML files. Dbafromthecold argues that \texttt{KubeDiagrams} addresses this challenge by generating accurate architecture diagrams directly from manifest files, allowing practitioners to visualize relationships between services, volumes, StatefulSets, and namespaces. The author concludes that \texttt{KubeDiagrams} significantly simplifies the task of documenting complex Kubernetes environments and provides immediate value to practitioners managing stateful applications. The blog author explicitly states that : \textit{“So having the ability to easily generate diagrams is really helpful… because we all should be documenting everything, right?”}; \textit{“It works really well and is a great way to visualise objects in Kubernetes.”}

\setlength{\tabcolsep}{4.5pt}
\begin{table}[ht]
\centering
\caption{Practitioner Statements about \texttt{KubeDiagrams} by theme and source}
\label{tab:quotes-summary}
\begin{tabular}{lccccccc}
\hline
\textbf{Theme} & \cite{saharan2025kubediagrams} & \cite{mrplanb2025kubediagrams} & \cite{dbafromthecold2025sql} & \cite{dailydev2025kubediagrams} & \cite{reddit2025kubediagrams020} & \cite{reddit2025kubediagramslucid} \\
\hline
Automation of generation & 4 & 2 & 2 & 1 & 1 & 0\\
\rowcolor{lightgray}Doc. maintenance & 3 & 2 & 1 & 1 & 1 & 0 \\
Usability and simplicity & 2 & 3 & 1 & 1 & 0 & 1 \\
\rowcolor{lightgray}Community appreciation & 1 & 1 & 2 & 0 & 2 & 1 \\
Cluster-wide/IaC support & 0 & 1 & 2 & 0 & 2 & 0  \\
\rowcolor{lightgray}Visual quality & 1 & 1 & 1 & 0 & 1 & 1 \\
Integration into workflows & 0 & 1 & 0 & 0 & 2 & 0 \\
\rowcolor{lightgray}Feature requests& 0 & 0 & 0 & 0 & 3 & 0 \\
\hline
\end{tabular}
\end{table}

\begin{figure}[ht]
  \centering
  \includegraphics[width=0.9\columnwidth]{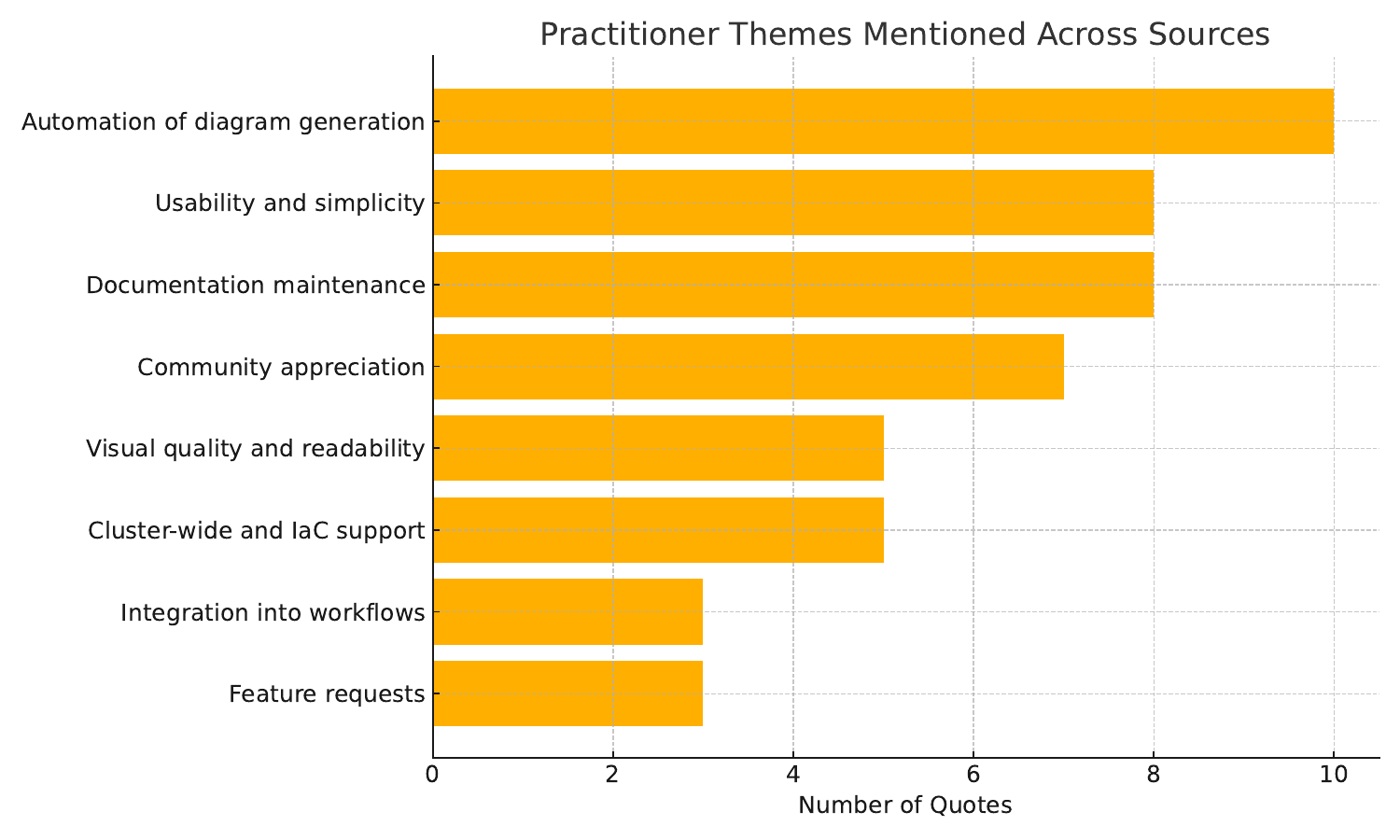}
  \caption{Frequency of practitioner themes related to \texttt{KubeDiagrams} across multiple sources.}
  \label{fig:kubediagrams_themes}
\end{figure}

Table~\ref{tab:quotes-summary} and Fig.~\ref{fig:kubediagrams_themes} highlight recurring themes in practitioner feedback about \texttt{KubeDiagrams} from the six sources, drawn from multiple blog posts, community threads, and informal reviews. The most prominent theme is automation of diagram generation (10 mentions), which reflects a widespread desire to replace manual, error-prone documentation with tools that generate architecture views directly from live Kubernetes state or configuration artifacts. 

Closely following are documentation maintenance, usability and simplicity, each mentioned 8 times, underscoring the tool’s appeal as both a practical aid for maintaining up-to-date system knowledge and a lightweight addition to existing workflows. Themes such as community appreciation (7 mentions) and visual quality and readability (5 mentions) confirm a positive reception regarding the aesthetics and impact of the diagrams, especially when compared to traditional tools like Lucidchart. 

Less frequent but still significant are mentions of cluster-wide and IaC support, integration into workflows (e.g., GitHub Actions, serverless deployments), and feature requests (e.g., HTML outputs, filtering ReplicaSets), indicating areas where users see potential for future development.

Overall, the distribution of feedback confirms that \textbf{\texttt{KubeDiagrams} addresses concrete pain points in Kubernetes-based DevOps practices and is valued for its automation, clarity, and low-friction integration into real-world tooling environments}.

\section{Limitations}
\label{sec:limitations}

Although \texttt{KubeDiagrams} supports a wide range of Kubernetes resource types (including CRDs), it does not visualize ephemeral system components such as \texttt{Events}, \texttt{TokenReview}, or real-time metrics. Furthermore, observability and runtime introspection remain outside its scope.

\texttt{KubeDiagrams} currently prioritizes static visualization and automation over interactivity. The tool generates architecture diagrams as static images or DOT files, which limits user exploration during runtime analysis. Engineers who require zoomable, clickable, or searchable diagrams must integrate \texttt{KubeDiagrams} outputs into external visualization platforms or manually post-process the results.

The tool also depends heavily on Kubernetes label conventions to infer application structure. When users omit or inconsistently apply labels such as \texttt{app}, \texttt{tier}, or \texttt{component}, the resulting diagrams may flatten logical groupings or misrepresent boundaries between services. While configuration files allow partial control over layout, \texttt{KubeDiagrams} does not validate semantic intent beyond syntactic matching.

The rendering engine relies on Graphviz for layout and output generation. This decision simplifies integration but constrains layout flexibility, especially for very large or deeply nested deployments. Developers cannot fine-tune spacing, alignment, or edge routing through the command line, which may frustrate users who expect WYSIWYG control or web-based refinement.

Finally, the project has not undergone a formal usability study. The authors collected community feedback from blogs and social media, but did not conduct structured evaluations with real-world teams. As a result, the extent to which \texttt{KubeDiagrams} improves onboarding speed, debugging accuracy, or architectural decision-making remains an open question.

\section{Conclusion}
\label{sec:conclusion}

\texttt{KubeDiagrams} addresses a practical gap in Kubernetes-based system development: the absence of reliable, up-to-date architectural diagrams. The tool transforms infrastructure-as-code artifacts and live cluster states into meaningful visual representations, helping teams understand, document, and communicate complex deployments without manual overhead.

By supporting a broad set of Kubernetes resource types and offering customization through declarative configuration, \texttt{KubeDiagrams} adapts to real-world use cases across development, operations, and compliance. Its design favors automation and scriptability, which makes it easy to integrate into CI/CD pipelines and documentation workflows.

Practitioners have already adopted the tool to improve onboarding, incident response, and environment comparison. Their unsolicited feedback confirms the value of real-time diagram generation grounded in actual system state. Thus, this work demonstrates that simple, scriptable tools can deliver meaningful improvements in DevOps practices when they align closely with platform semantics and user workflows. As Kubernetes continues to grow in complexity and scale, maintaining accurate architectural views will remain critical. \texttt{KubeDiagrams} offers a step toward that goal by turning system structure into a first-class, reproducible artifact.

\texttt{KubeDiagrams} currently produces static diagrams work well for documentation, but fall short in exploratory analysis. A next step involves adding interactive capabilities, such as zooming, filtering, and drill-down views. These features would help users navigate large diagrams more effectively and isolate specific components during debugging or audits. Furthermore, the tool lacks support for runtime observability data. By integrating metrics and logs from systems like Prometheus or Fluent Bit, future versions could offer hybrid views that combine deployment structure with live system behavior. This would turn \texttt{KubeDiagrams} from a static visualizer into a dynamic systems lens.

Finally, while the current visual encoding handles core and custom Kubernetes resources, users still face limitations when visualizing certain control-plane entities, ephemeral objects, or horizontal platform layers, \textit{e.g.}, service meshes. Expanding support for these advanced features requires better parsing logic and new iconography. Another priority involves scaling the layout engine for very large clusters. Some users have reported cluttered or unreadable diagrams when visualizing dozens of namespaces or hundreds of resources. Exploring alternative layout algorithms or integrating incremental rendering could improve performance and clarity. 

%Finally, the project would benefit from a structured usability study. Practitioners have shown interest through blogs and social media, but controlled experiments could uncover gaps, validate assumptions, and guide refinements. Engaging with teams in industry and academia will help shape the roadmap around actual workflow needs.

\section*{Acknowledgment}

This work was supported by two French government grants managed by the Agence Nationale de la Recherche under the France 2030 program, reference ``ANR-22-CE25-0009" and ``ANR-23-PECL-0008", and Natural Sciences and Engineering Research Council of Canada Discovery Grant program, reference RGPIN-2019-05339.

%\newpage

\balance
\bibliographystyle{IEEEtran}
\bibliography{references}

% Generated by IEEEtran.bst, version: 1.14 (2015/08/26)
\begin{thebibliography}{10}
\providecommand{\url}[1]{#1}
\csname url@samestyle\endcsname
\providecommand{\newblock}{\relax}
\providecommand{\bibinfo}[2]{#2}
\providecommand{\BIBentrySTDinterwordspacing}{\spaceskip=0pt\relax}
\providecommand{\BIBentryALTinterwordstretchfactor}{4}
\providecommand{\BIBentryALTinterwordspacing}{\spaceskip=\fontdimen2\font plus
\BIBentryALTinterwordstretchfactor\fontdimen3\font minus
  \fontdimen4\font\relax}
\providecommand{\BIBforeignlanguage}[2]{{%
\expandafter\ifx\csname l@#1\endcsname\relax
\typeout{** WARNING: IEEEtran.bst: No hyphenation pattern has been}%
\typeout{** loaded for the language `#1'. Using the pattern for}%
\typeout{** the default language instead.}%
\else
\language=\csname l@#1\endcsname
\fi
#2}}
\providecommand{\BIBdecl}{\relax}
\BIBdecl

\bibitem{10.1007/3-540-45848-4_57}
J.~Ellson, E.~Gansner, L.~Koutsofios, S.~C. North, and G.~Woodhull,
  ``Graphviz--- open source graph drawing tools,'' in \emph{Graph Drawing},
  P.~Mutzel, M.~J{\"u}nger, and S.~Leipert, Eds.\hskip 1em plus 0.5em minus
  0.4em\relax Berlin, Heidelberg: Springer Berlin Heidelberg, 2002, pp.
  483--484.

\bibitem{kubeview}
B.~Coleman, ``Kubeview: Kubernetes cluster visualiser and graphical explorer,''
  \url{https://github.com/benc-uk/kubeview}, 2019, accessed: 2025-05-22.

\bibitem{kubediagrams}
P.~Merle, ``{KubeDiagrams: A Kubernetes architecture visualization tool},''
  \url{https://github.com/philippemerle/KubeDiagrams}, 2024, accessed:
  2025-05-22.

\bibitem{k8sviz}
M.~Kimura, ``k8sviz: Kubernetes visualization tool,''
  \url{https://github.com/mkimuram/k8sviz}, 2019, accessed: 2025-05-22.

\bibitem{k8sdiagrams2021}
trois six, ``k8s-diagrams: Generate diagrams from kubernetes yaml manifests,''
  \url{https://github.com/trois-six/k8s-diagrams}, 2021, accessed: 2025-05-22.

\bibitem{grucloud}
\BIBentryALTinterwordspacing
GruCloud, ``Grucloud: Iac diagram and deployment tool,''
  \url{https://github.com/grucloud/grucloud}, 2020, accessed: 2025-05-22.
  [Online]. Available: \url{https://github.com/grucloud/grucloud}
\BIBentrySTDinterwordspacing

\bibitem{kocierik2024k8sdiagram}
M.~Kocierik, ``k8s-to-diagram: Kubernetes architecture diagram generator,''
  \url{https://github.com/kocierik/k8s-to-diagram}, 2024, accessed: 2025-05-22.

\bibitem{reactk8sviewer}
{SocialGouv}, ``{react-k8s-viewer: Kubernetes UI components in React},''
  \url{https://github.com/SocialGouv/react-k8s-viewer}, 2021, accessed:
  2025-05-22.

\bibitem{k8sdiagrampreviewer}
J.~Mills, ``K8s diagram previewer,''
  \url{https://github.com/jimmymills/k8s-diagram-previewer}, 2021, accessed:
  2025-05-22.

\bibitem{k8s-diagrams}
J.~Ángel, ``k8s-diagrams: Generate diagrams from kubernetes manifests,''
  \url{https://github.com/imjoseangel/k8s-diagrams}, 2021, accessed:
  2025-05-22.

\bibitem{kube-diagram}
K.~A. Howell, ``kube-diagram: Generate diagrams from kubernetes yaml
  manifests,'' \url{https://github.com/kahowell/kube-diagram}, 2022, accessed:
  2025-05-22.

\bibitem{kubeDiagrams2020}
\BIBentryALTinterwordspacing
Sunny, ``kube-diagrams: Generate architecture diagrams for kubernetes
  resources,'' \url{https://github.com/sunny10031982/kube-diagrams}, 2020,
  accessed: 2025-05-22. [Online]. Available:
  \url{https://github.com/sunny10031982/kube-diagrams}
\BIBentrySTDinterwordspacing

\bibitem{k8d}
N.~Schleicher, ``k8d: Kubernetes diagrams from yaml files,''
  \url{https://github.com/NickSchleicher/k8d}, 2020, accessed: 2025-05-22.

\bibitem{k8s_diagram2024}
MrSir, ``{k8s\_diagram}: A kubernetes architecture diagram generator,''
  \url{https://github.com/MrSir/k8s_diagram}, 2024, accessed: 2025-05-22.

\bibitem{kubedraw2024}
B.~Nam, ``Kubedraw: A kubernetes architecture visualization tool,''
  \url{https://github.com/B0nam/kubedraw}, 2024, accessed: 2025-05-22.

\bibitem{saharan2025kubediagrams}
\BIBentryALTinterwordspacing
A.~Saharan, ``Generate kubernetes architecture maps directly from your
  cluster,'' Mar. 2025, accessed: 2025-05-21. [Online]. Available:
  \url{https://blog.abhimanyu-saharan.com/posts/generate-kubernetes-architecture-maps-directly-from-your-cluster}
\BIBentrySTDinterwordspacing

\bibitem{mrplanb2025kubediagrams}
\BIBentryALTinterwordspacing
Mr.PlanB, ``Kubediagrams 0.2.0 makes it way easier to visualize your kubernetes
  setup,'' Mar. 2025, accessed: 2025-05-21. [Online]. Available:
  \url{https://medium.com/weeklycloud/kubediagrams}
\BIBentrySTDinterwordspacing

\bibitem{dbafromthecold2025sql}
\BIBentryALTinterwordspacing
dbafromthecold, ``Visualising sql server in kubernetes,'' Feb. 2025, accessed:
  2025-05-21. [Online]. Available:
  \url{https://dbafromthecold.com/2025/02/06/visualising-sql-server-in-kubernetes/}
\BIBentrySTDinterwordspacing

\bibitem{dailydev2025kubediagrams}
\BIBentryALTinterwordspacing
{DailyOpenSourceTools}, ``Kubediagrams: Generate live kubernetes architecture
  maps effortlessly,'' mar 2025, accessed: 2025-05-21. [Online]. Available:
  \url{https://app.daily.dev/posts/kubediagrams-e35zcloui}
\BIBentrySTDinterwordspacing

\bibitem{reddit2025kubediagrams020}
\BIBentryALTinterwordspacing
{Reddit users}, ``{KubeDiagrams 0.2.0 is out!}'' Mar. 2025, accessed:
  2025-05-21. [Online]. Available:
  \url{https://www.reddit.com/r/kubernetes/comments/1jjjw6j/kubediagrams_020_is_out/}
\BIBentrySTDinterwordspacing

\bibitem{reddit2025kubediagramslucid}
\BIBentryALTinterwordspacing
{u/Anonymous}, ``Kubediagrams,'' Mar. 2025, accessed: 2025-05-21. [Online].
  Available:
  \url{https://www.reddit.com/r/kubernetes/comments/1ihjujy/kubediagrams/}
\BIBentrySTDinterwordspacing

\end{thebibliography}

\end{document}